\documentclass[acmsmall]{acmart}

\usepackage{metalogo}
\usepackage{smartdiagram}
\usepackage{forest}
\usepackage{cleveref}
\usepackage{booktabs}
\usepackage{multirow}
\usepackage{makecell}
\usepackage{subfigure}
\usepackage[misc]{ifsym} 
\usepackage{balance}
\usepackage{listings}
\usepackage{color, xcolor}
\usepackage{tcolorbox}
\usepackage{amsmath}
\usepackage{algorithm}
\usepackage{graphicx}
\usepackage{algorithmicx}
\usepackage{algpseudocode}
\usepackage{textcomp}
\usepackage{diagbox}
\usepackage{colortbl}

\setcopyright{acmlicensed}
\copyrightyear{2025}
\acmYear{2025}
\acmDOI{XXXXXXX.XXXXXXX}

\acmJournal{JACM}
\acmVolume{37}
\acmNumber{4}
\acmArticle{111}
\acmMonth{8}

\begin{document}

\title{LogDB: Multivariate Log-based Failure Diagnosis for Distributed Databases (Extended from MultiLog)}

\author{Lingzhe Zhang}
\affiliation{%
	\institution{Peking University}
	\city{Beijing}
	\country{China}}
\email{zhang.lingzhe@stu.pku.edu.cn}

\author{Tong Jia$^{\ast}$}
\thanks{$^{\ast}$Corresponding author}
\affiliation{%
	\institution{Peking University}
	\city{Beijing}
	\country{China}}
\email{jia.tong@pku.edu.cn}

\author{Mengxi Jia}
\affiliation{%
	\institution{Peking University}
	\city{Beijing}
	\country{China}}
\email{mxjia@pku.edu.cn}

\author{Ying Li$^{\ast}$}
\affiliation{%
	\institution{Peking University}
	\city{Beijing}
	\country{China}}
\email{li.ying@pku.edu.cn}

\renewcommand{\shortauthors}{Lingzhe Zhang et al.}

\begin{abstract}
	Distributed databases, as the core infrastructure software for internet applications, play a critical role in modern cloud services. However, existing distributed databases frequently experience system failures and performance degradation, often leading to significant economic losses. Log data, naturally generated within systems, can effectively reflect internal system states. In practice, operators often manually inspect logs to monitor system behavior and diagnose anomalies, a process that is labor-intensive and costly. Although various log-based failure diagnosis methods have been proposed, they are generally not tailored for database systems and fail to fully exploit the internal characteristics and distributed nature of these systems. To address this gap, we propose LogDB, a log-based failure diagnosis method specifically designed for distributed databases. LogDB extracts and compresses log features at each database node and then aggregates these features at the master node to diagnose cluster-wide anomalies. Experiments conducted on the open-source distributed database system Apache IoTDB demonstrate that LogDB achieves robust failure diagnosis performance across different workloads and a variety of anomaly types.
\end{abstract}




\maketitle

\section{Introduction}

With the rapid development of artificial intelligence technologies, AIOps has been widely adopted for IT operations in software systems~\cite{dang2019aiops}. AIOps analyzes large-scale data collected from various operational tools and devices using machine learning and other algorithms, enabling the automatic detection and real-time response to incidents, thereby enhancing the capability and automation level of IT operations. Since system logs record the states of running processes and critical events in fine detail, they serve as a rich data source for failure diagnosis. As a result, under the trend of AIOps, automatic failure diagnosis technologies based on system logs have developed rapidly for large-scale distributed systems~\cite{sui2023logkg,lin2016log,yuan2019approach, jia2022augmenting,jia2021logflash,du2017deeplog,meng2019loganomaly,zhang2019robust,yang2021semi,liu2022uniparser, zhang2024towards, zhang2025scalalog, zhang2025xraglog, zhang2025agentfm, zhang2025thinkfl, zhang2024reducing}. Log-based failure diagnosis techniques typically analyze historical system logs and corresponding anomaly labels during the training phase to construct models that capture request execution paths under different anomalies. During runtime, they perform real-time matching using logs to achieve accurate fault classification and diagnosis~\cite{zhang2024survey}.

Distributed databases, as the fundamental infrastructure software for internet applications, play a critical role in modern cloud service products. In the context of the current wave of domestic software development, a series of distributed databases have emerged, such as Alibaba OceanBase~\cite{yang2022oceanbase}, PingCAP TiDB~\cite{huang2020tidb}, and Apache IoTDB~\cite{wang2020apache}. However, distributed databases frequently experience system failures and performance degradation, which often result in substantial financial losses. For instance, Alibaba Cloud reported that intermittent slow queries (iSQs) alone can lead to billions of dollars in annual losses~\cite{ma2020diagnosing}. Similarly, Amazon reported that every additional 0.1 seconds of database-induced latency could result in approximately 1\% financial loss. Therefore, diagnosing anomalies in distributed databases and mitigating their impacts is of critical importance.

This study focuses on the failure diagnosis problem for distributed databases based on software log data. Existing failure diagnosis methods for databases can be categorized into: (1) diagnosis of overall database operational states, (2) diagnosis targeting SQL queries, and (3) diagnosis focusing on internal database components. However, most of these methods are based on monitoring metrics rather than log data, and there is still a lack of approaches specifically designed for database software logs. In the microservices domain, two main categories of log-based failure diagnosis methods exist: graph-based algorithms and deep learning-based algorithms. Graph-based methods, such as LogKG~\cite{sui2023logkg} and LogCluster~\cite{lin2016log}, combine clustering techniques and knowledge graphs to classify fault types. Deep learning-based methods, such as Cloud19~\cite{yuan2019approach}, convert logs into vector representations using Word2Vec and apply neural networks to classify anomalies. Although these log-based methods have achieved excellent results in microservices scenarios, several challenges arise when applying them to distributed databases:

\begin{itemize}
	\item Existing failure diagnosis algorithms are designed for single-node systems and do not consider the multi-node, multi-replica characteristics of distributed databases. In distributed databases, to ensure data consistency, cluster nodes are classified into leaders and followers, implying that different nodes contain different information;
	\item The correspondence between system logs and distributed database anomalies is complex. For performance reasons, not every write or query request generates corresponding logs, making it challenging to extract useful information from incomplete system logs;
	\item Due to the complexity of distributed database architectures, the types of anomalies are richer compared to those in microservices scenarios. As a result, applying current algorithms to distributed databases leads to low diagnostic accuracy.
\end{itemize}

To address the above challenges, we propose \textbf{LogDB}, a log-based failure diagnosis framework for distributed databases that extracts and compresses log features at each database node and aggregates them at the master node to diagnose cluster-wide anomalies. LogDB is a failure diagnosis method extended from MultiLog~\cite{zhang2024multivariate}. Specifically, we first collect sequence patterns, event counts, and semantic information from the logs of each independent node within the database cluster. We then employ a Long Short-Term Memory (LSTM) network enhanced with a self-attention mechanism to encode and predict these features, generating an anomaly feature matrix for each node. Finally, we construct a cluster-level classifier that uses a Variational Autoencoder (VAE) to extract implicit representations from the anomaly matrices across nodes, followed by a meta-classifier to accurately identify the anomaly types affecting the entire cluster.

We conduct extensive experiments on the open-source distributed database system Apache IoTDB, demonstrating that LogDB consistently achieves superior failure diagnosis performance across a wide range of workloads and fault scenarios.

In summary, our main contributions are as follows:
\begin{itemize}
	\item We propose \textbf{LogDB}, a multivariate log-based failure diagnosis framework for distributed databases, which effectively integrates logs from multiple nodes to diagnose anomalies at the cluster level.
	\item We design a feature extraction method that captures sequence, count, and semantic information from logs using an LSTM network with self-attention, enabling more informative node-level representations.
	\item We introduce a cluster-level anomaly classification approach that leverages a Variational Autoencoder (VAE) to fuse multi-node anomaly matrices and extract implicit features for accurate and robust failure diagnosis.
	\item We validate LogDB through comprehensive experiments on Apache IoTDB, showing that it achieves excellent anomaly detection performance under diverse workloads and multiple types of system failures.
\end{itemize}

\section{Problem Definition}

\begin{figure*}[htbp]
	\centering
	\includegraphics[width=1\linewidth]{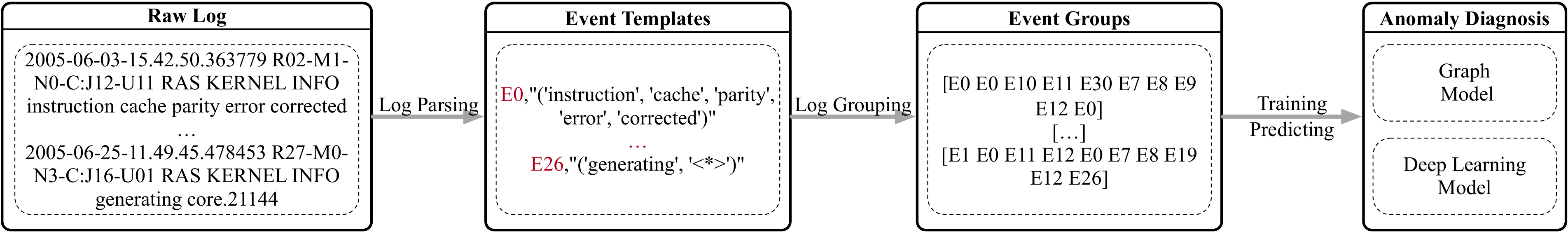}
	\caption{Common Workflow of Log-based failure diagnosis}
	\label{fig:common-workflow}
\end{figure*}

The goal of failure diagnosis is to automatically identify the root causes of system anomalies, thereby assisting operators in promptly locating and resolving issues. Based on diagnostic objectives and outputs, existing log-based failure diagnosis methods can be categorized into five types: execution path analysis, failure-related log extraction, specific root cause analysis, diagnosis with auxiliary data, and failure type identification. This work focuses on the problem of \textbf{failure type identification}, which is formally defined as follows.

\subsection{Failure Diagnosis}

Failure type identification can be formulated as a multi-class classification problem. Given system runtime data, the task is to analyze detected anomalous periods and accurately classify them into predefined failure types~\cite{sui2023logkg, lin2016log, yuan2019approach}. Specifically, the process first involves identifying and extracting anomalous patterns and features from the data, then mapping these features to corresponding failure type labels via a trained model. Through this approach, the system can automatically determine the type of failure associated with an anomaly, providing targeted guidance for issue resolution and helping to maintain stable system operations.

Formally, given an anomalous data segment $D$, we first apply a feature extraction method $F$ to obtain a feature representation $DF$, as defined in Equation~\ref{eq:df}:

\begin{equation}
	DF = F(D)
	\label{eq:df}
\end{equation}

Based on the extracted feature $DF$, the goal is to map it to one of the predefined failure type labels $label_i$ in the label set $L$. This mapping process, conducted via a trained model $M$, can be expressed as Equation.~\ref{eq:label}, where $label_i \in L$.

\begin{equation}
	M(DF) \rightarrow label_i
	\label{eq:label}
\end{equation}

\subsection{Log Processing}
\label{sec:log-process}

System logs, naturally generated during the operation of distributed databases, record rich internal states, including database queries, transaction executions, system errors, and node heartbeats. To extract valuable information from these complex logs, state-of-the-art methods generally follow the workflow shown in Fig.~\ref{fig:common-workflow}, which includes three main stages: log parsing, log grouping, and model training and prediction~\cite{sui2023logkg,lin2016log,yuan2019approach, jia2022augmenting,jia2021logflash,du2017deeplog,meng2019loganomaly,zhang2019robust,yang2021semi,liu2022uniparser}.

Raw database logs are semi-structured data, containing both fixed fields (e.g., timestamps, severity levels) and dynamic content. To facilitate downstream tasks, log parsing transforms each log message into an event template that captures the invariant part of the message along with its associated variable parameters. For example, from the log message shown in Fig.~\ref{fig:common-workflow}, ``2005-06-03-15.42.50.363779 R02-M1-N0-C:J12-U11 RAS KERNEL INFO instruction cache parity error corrected," the corresponding event template can be extracted as ``E0, ('instruction', 'cache', 'parity', 'error', 'corrected').'' Since each log corresponds to a specific log print statement in the source code, multiple log messages may map to the same event template. After parsing, an individual event template is denoted as $e_i$, and the complete set of templates is denoted as $w = \{e_1, e_2, ..., e_n\}$.

After parsing, the logs are organized into sequences through log grouping algorithms, typically based on fixed-size windows. A grouped log sequence can be represented as $E = (e_{(s_1)}, e_{(s_2)}, ..., e_{(s_N)})$, where $s_i$ indexes the parsed log messages and $N$ is the window size. In the subsequent failure diagnosis phase, each grouped sequence $E$ serves as a basic diagnostic unit, and the task becomes to classify $E$ into one of the failure type labels, that is, to perform $M(E) \rightarrow label_i$.

\section{LogDB}

\begin{figure}[htbp]
	\centering
	\includegraphics[width=1\linewidth]{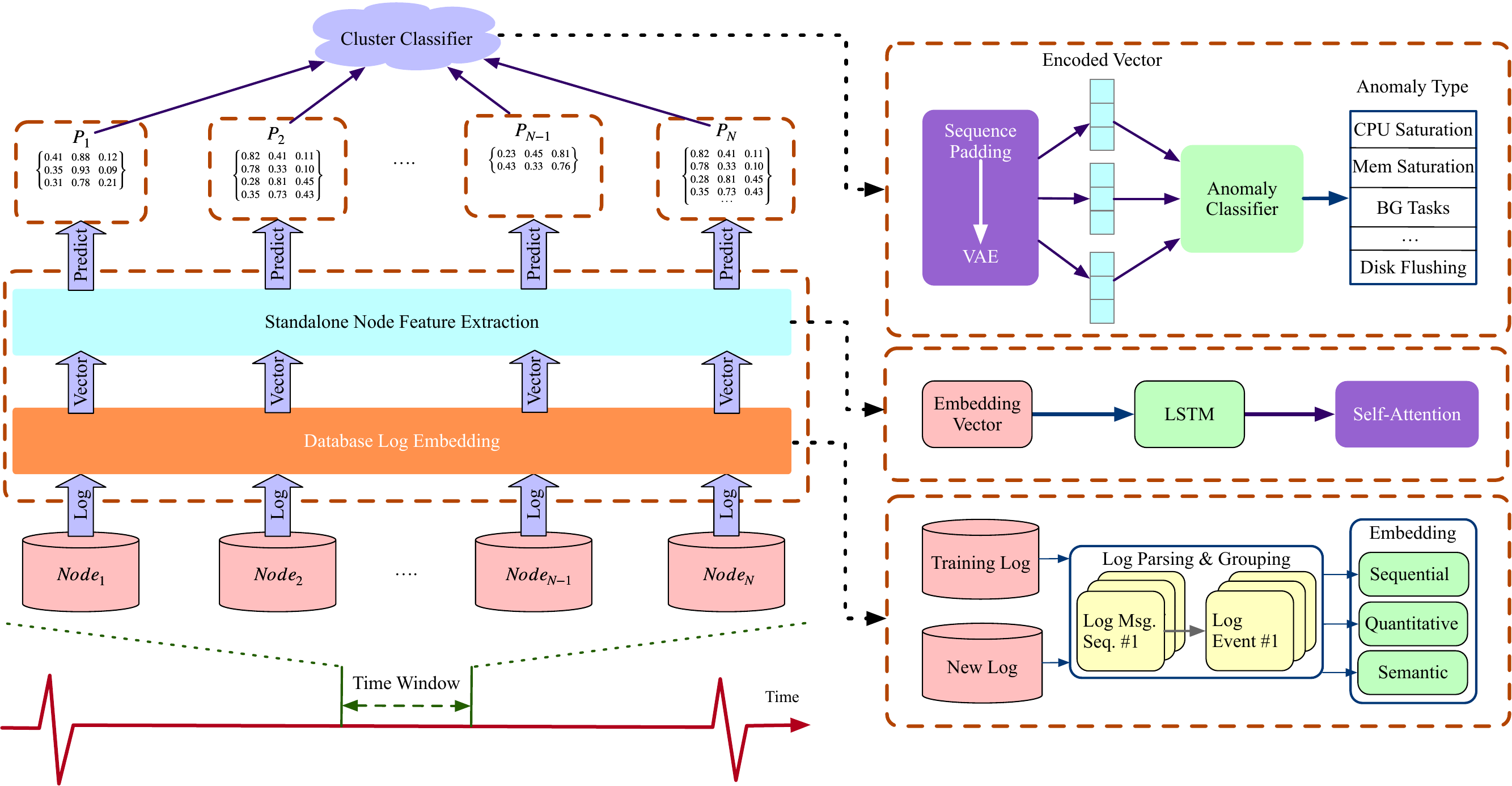}
	\caption{Architecture of LogDB}
	\label{fig:logdb}
\end{figure}

The overall workflow and architecture of the proposed method are illustrated in Fig.~\ref{fig:logdb}. LogDB is a log-based failure diagnosis method specifically designed for distributed databases. It comprises three key modules: Database Log Embedding, Standalone Node Feature Extraction based on self-attention mechanisms, and Cluster Anomaly Classification based on Variational Autoencoders (VAEs).

The Log Embedding module first transforms semi-structured log data into structured vector representations. The Feature Extraction module compresses log data from multiple nodes using a Long Short-Term Memory (LSTM) network enhanced with a Self-Attention mechanism to extract critical feature representations. Finally, the Cluster Anomaly Classification module aggregates the extracted features across nodes using Sequence Padding and a Variational Autoencoder (VAE)~\cite{kusner2017grammar}, and determines the anomaly type through an Anomaly Classifier.

\subsection{Database Log Embedding}

Building on existing log processing techniques, LogDB further tailors log embeddings to database-specific anomaly features by converting raw logs into feature vectors. Specifically, logs are divided into fixed-size time windows $T$, and each log sequence within a window $t$ is denoted as $S_t = (s_1, s_2, \ldots, s_{N_t})$. The preprocessing techniques described in Section~\ref{sec:log-process}, including log parsing and grouping, are then applied.

For each log group $S_t = (s_1, s_2, \ldots, s_N)$, three types of embeddings are performed: Sequential Embedding, Quantitative Embedding, and Semantic Embedding.

\textbf{Sequential Embedding} organizes parsed log events into a natural sequence pattern. Each event template is denoted as $e_i$, and the complete set of templates across the distributed database is $\omega = \{e_1, e_2, \ldots, e_n\}$. After grouping, the log group can be expressed as Equation~\ref{eq:log-group}.

\begin{equation}
	E = (e_{(s1)}, e_{(s2)}, \ldots, e_{(sN)})
	\label{eq:log-group}
\end{equation}

\textbf{Quantitative Embedding} captures the frequency-based patterns in the log sequences. Under normal operations, logs often maintain specific quantitative relationships, such as matching numbers of "create Memtable" and "flush Memtable to file" events in a database. Disruptions to these patterns may indicate anomalies. For a sequence $S = (s_1, s_2, \ldots, s_N)$, the count of each event template $e_k$ is aggregated into a frequency vector, as illustrated in Equation~\ref{eq:vector}.

\begin{equation}
	C = (c_{(e1)}, c_{(e2)}, \ldots, c_{(eN)})
	\label{eq:vector}
\end{equation}

\textbf{Semantic Embedding} extracts semantic information from log event templates. Since templates often contain meaningful English words, LogDB preprocesses templates by removing non-word characters and splitting compound words (e.g., "NullPointerException") using camel case segmentation. Then, it applies a pre-trained FastText model based on the Common Crawl Corpus to obtain word embeddings. Each word $w$ is assigned a TF-IDF weight as illustrated in Equation~\ref{eq:tf} and Equation~\ref{eq:idf}.

\begin{equation}
	TF(w) = \frac{\#w}{\#W}
	\label{eq:tf}
\end{equation}

\begin{equation}
	IDF(w) = \log\left(\frac{\#L}{\#L_w}\right)
	\label{eq:idf}
\end{equation}

where $\#w$ is the frequency of word $w$ in a template, $\#W$ is the total number of words, $\#L$ is the total number of templates, and $\#L_w$ is the number of templates containing $w$. Each word vector is weighted by its TF-IDF score, and the semantic embedding vector $V$ for a log template is computed as Equation~\ref{eq:tf-idf}.

\begin{equation}
	V = \frac{1}{W} \sum_{i=1}^{W} \epsilon_i
	\label{eq:tf-idf}
\end{equation}

\subsection{Standalone Node Feature Extraction}

For each embedding vector—$E$, $C$, and $V$—LogDB uses an LSTM enhanced with a Self-Attention mechanism to extract salient features. The LSTM outputs a sequence $H = [h_1, h_2, \ldots, h_N]$, where $h_n$ is the hidden state at position $n$.

The attention score for each hidden state is computed as Equation~\ref{eq:score}.

\begin{equation}
	\alpha_{n} = \frac{\exp(\text{score}(h_n, h_N))}{\sum_{n'=1}^{N} \exp(\text{score}(h_{n'}, h_N))}
	\label{eq:score}
\end{equation}

where the scoring function is defined as $\text{score}(h_m, h_M) = h_m^\intercal W h_M$, with $W$ being a learnable weight matrix. The context vector $c$ is the weighted sum of hidden states as Equation~\ref{eq:c}.

\begin{equation}
	c = \sum_{m=1}^{M} \alpha_m h_m
	\label{eq:c}
\end{equation}

Each enhanced embedding $EC$ is obtained by concatenating $c$ and the final hidden state $h_M$: $EC = [c; h_M]$. Finally, all enhanced vectors are concatenated and passed through a fully connected layer as Equation~\ref{eq:p}

\begin{equation}
	p = FC([EC_E; EC_C; EC_V])
	\label{eq:p}
\end{equation}

\subsection{Cluster Anomaly Classification}

After extracting the log features from each database node, LogDB employs a distributed cluster anomaly classifier based on a variational autoencoder (VAE) to diagnose the type of cluster anomalies. To standardize the feature length output by each node, LogDB first applies a sequence padding method. Then, to unify and compress the padded feature vectors, LogDB utilizes a VAE to extract the latent features of the log data. Finally, an anomaly classifier is used to aggregate these standardized latent features and classify the anomaly within the current time window.

Considering that different nodes may generate varying amounts of logs within the same time window $t$, as previously described, we regroup the logs within a time window into fixed-length segments of size $N$, and generate a feature vector $p$ for each group, where $||p||$ equals the predefined number of anomaly types. Therefore, for a database with $Q$ nodes, the log feature matrix of node $Node_i$ within each time window can be represented as Equation~\ref{eq: p_i}, where $p_{j}^{i}$ denotes the feature vector of the $j$-th group of logs from node $Node_i$, and $k_i$ denotes the number of groups for node $Node_i$.

\begin{equation}
	P_{i}=[p_{1}^{i},p_{2}^{i},...,p_{k_{i}}^{i}]
	\label{eq: p_i}
\end{equation}

To ensure a consistent size of the log feature matrices across database nodes, we perform sequence padding on each node's log feature matrix list. Specifically, we pad or truncate $P_i$ to a fixed sequence length $\beta$, as shown in Equation~\ref{eq: p_i'}. Here, $0_{||p||}=[0, 0, \dots, 0]$ represents a zero vector of length $||p||$, $\text{len}(P_i)$ denotes the length of $P_i$, and $P_{i}[0:\beta]$ refers to truncating $P_i$ to retain only the first $\beta$ elements. Through this operation, each node’s feature vectors are adjusted to the same length $\beta$, facilitating subsequent VAE processing.

\begin{equation}
	P_{i}^{'} =\begin{cases}
		P_{i} + 0_{||p||} \cdot (\beta - \text{len}(P_{i})), & \text{if } \text{len}(P_{i}) < \beta \\
		P_{i}[0:\beta], & \text{if } \text{len}(P_{i}) > \beta
	\end{cases}
	\label{eq: p_i'}
\end{equation}

Next, we apply a VAE to $P_{i}^{'}$ to unify and compress the padded feature vectors and to extract the latent representations of the log data. Specifically, the VAE consists of an encoder ($f_{enc}$), a latent space sampler, and a decoder ($f_{dec}$). As shown in Equation~\ref{eq: Z_i}, LogDB first applies the encoder function $f_{enc}$ to map $P_{i}^{'}$ into a latent representation $Z_i$ of fixed size $\theta$. In our implementation, $f_{enc}$ consists of three linear layers with ReLU activations. In addition to the latent output $Z_i$, the encoder also produces the mean $\mu_i$ and standard deviation $\sigma_i$ of the latent variables.

Subsequently, reparameterization is performed by sampling from the latent space, as shown in Equation~\ref{eq: z_i}, where $\epsilon_i$ is randomly sampled from a standard normal distribution. The decoder function $f_{dec}$ then reconstructs the original feature matrix from the sampled latent variable, as shown in Equation~\ref{eq: P_i^}. In our implementation, $f_{dec}$ consists of a linear layer with ReLU activation followed by a linear layer with Sigmoid activation. The overall loss function is the sum of mean squared error (MSELoss) and Kullback-Leibler Divergence, which is minimized during training.

\begin{equation}
	Z_{i} = f_{enc} (P_{i})
	\label{eq: Z_i}
\end{equation}

\begin{equation}
	z_{i}=\mu_{i}+\sigma_{i} \cdot \epsilon_{i}
	\label{eq: z_i}
\end{equation}

\begin{equation}
	\hat{P_{i}} = f_{dec} (Z_{i})
	\label{eq: P_i^}
\end{equation}

After compressing the feature matrices and obtaining the latent features, LogDB employs a cluster anomaly classifier to classify the anomaly type within the corresponding time window. The classifier first concatenates the latent features from all nodes to form a cluster feature matrix $Z$, as shown in Equation~\ref{eq: Z}. This cluster feature matrix $Z$ is then fed into a convolutional neural network (CNN), denoted as $f_{anomaly}$. Finally, the CNN outputs the probability distribution over different anomaly classes, as shown in Equation~\ref{eq: P_class}, where $P\{class_1, class_2, \dots, class_{||p||}\}$ represents the probability of each anomaly class, and $||p||$ is the total number of anomaly classes.

\begin{equation}
	Z = [Z_{1};Z_{2};...;Z_{N}]
	\label{eq: Z}
\end{equation}

\begin{equation}
	P\{class_1, class_2,...,class_{||p||}\} = f_{anomaly} (Z)
	\label{eq: P_class}
\end{equation}

\section{Experiment and Evaluation}

\subsection{Implementation and Benchmark}

We chose Apache IoTDB~\cite{wang2020apache} as our experimental platform for the following reasons: (1) it is a widely used open-source database; (2) it is a representative distributed database; (3) it has an active community.

All experiments were conducted using Apache IoTDB v1.2.2 running in a Docker environment. We deployed 4 Docker containers: 3 served as DataNodes and 1 as a ConfigNode. Each container was configured with 8 Intel(R) Xeon(R) Platinum 8260 CPUs running at 2.40GHz, 16GB of DIMM RAM, and a 1.1TB NVMe SSD, operating on OpenJDK 11.

For workload generation during database operations, we adopted TSBS~\cite{github:tsbs}, TPCx-IoT~\cite{poess2018analysis}, and IoT-Benchmark~\cite{liu2019benchmarking}. Specifically, TSBS is a widely used benchmark for evaluating time-series databases and is frequently referenced in database ranking platforms such as benchANT~\cite{seybold2021benchmarking}. TPCx-IoT is an industry-standard benchmark that provides verifiable metrics for performance, cost-effectiveness, and availability in commercial systems. IoT-Benchmark is a benchmarking tool provided by Apache IoTDB itself for comparing IoTDB with other time-series database solutions.

To simulate failures during database operation, we employed Chaos Mesh~\cite{bjornberg2021cloud}, an open-source cloud-native chaos engineering platform that supports a variety of fault injection types, enabling the simulation of real-world system failures. However, Chaos Mesh is limited to injecting system-level failures within Docker environments. For workload-related or internal configuration failures, we dynamically adjusted read/write loads or performed hot configuration changes.

To comprehensively cover common types of failures in distributed databases, we analyzed anomalies reported in microservices-oriented and database-oriented diagnostic research, in addition to drawing from practical engineering experience. We categorized failures into two main types: (1) system anomalies and (2) database anomalies.

\begingroup
\setlength{\tabcolsep}{15pt}
\begin{table}[htbp]
	\centering
	\caption{Injected Anomaly Types}
	\label{tab: anomaly-type}
	\begin{tabular}{ccc}
		\toprule
		No. & Anomaly & Type \\
		\midrule
		1 & CPU Saturation & System anomaly \\
		2 & Memory Saturation & System anomaly \\
		3 & Network Bandwidth Limited & System anomaly \\
		4 & Export Operations & Database anomaly \\
		5 & Import Operations & Database anomaly \\
		6 & Too Many Background Tasks & Database anomaly \\
		7 & Too Frequent Disk Flushing & Database anomaly \\
		\bottomrule
	\end{tabular}
\end{table}
\endgroup

As shown in Table~\ref{tab: anomaly-type}, we injected three system-level anomalies and four database-level anomalies. Among them, for the "Too Many Background Tasks" anomaly (No.6), we used Compaction tasks as a representative background task during injection, which are typical in LSM-tree based database systems~\cite{zhang2024time, kang2022separation}. Compaction tasks can exert significant pressure on CPU and memory resources depending on system configurations and may lead to issues such as write stalls. Meanwhile, the "Too Frequent Disk Flushing" anomaly (No.7) is a common issue across different types of databases, affecting the speed of memory-disk interaction and significantly impacting the frequency of file creation and overall system performance.

In terms of injection methods, anomalies No.1 to No.3 were injected using Chaos Mesh, anomalies No.4 and No.5 were injected through dynamic load adjustment, and anomalies No.6 and No.7 were injected via hot database configuration changes.

\begin{figure}[htbp]
	\centering
	\includegraphics[width=1\linewidth]{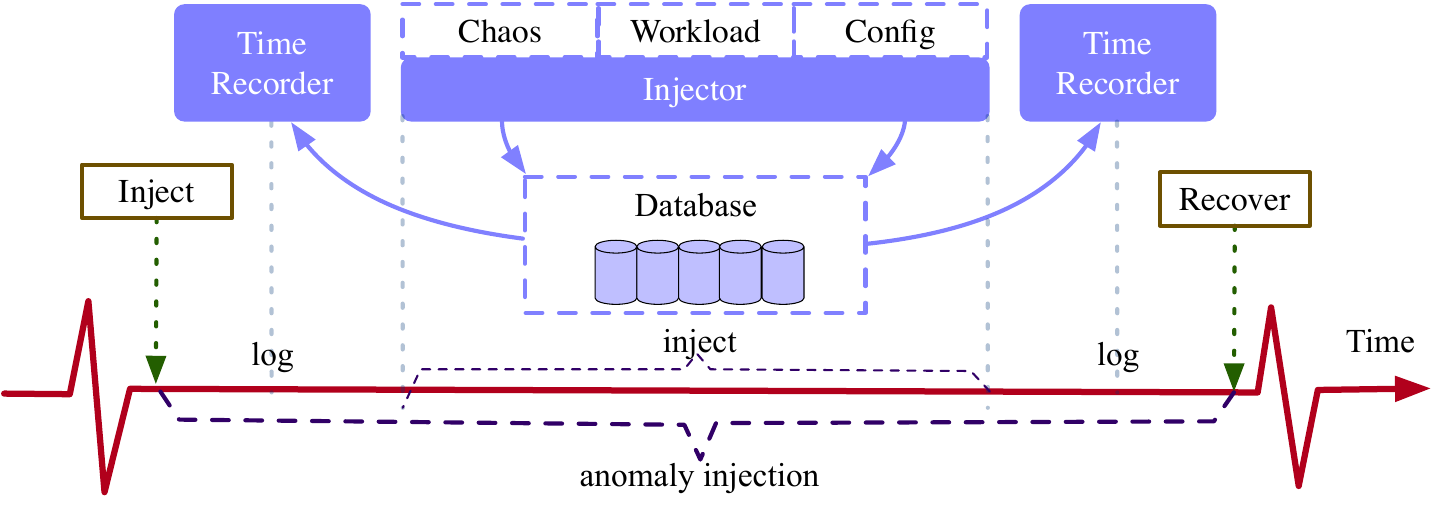}
	\caption{Anomaly Injection Procedure}
	\label{fig: anomaly-injection}
\end{figure}

The overall anomaly injection process is illustrated in Figure~\ref{fig: anomaly-injection}. Initially, the database operates normally for a certain period. Then, an anomaly is randomly selected and injected into a specific node of the distributed database. Before each injection, a timestamp is recorded, and the anomaly is continuously injected for a fixed duration. Afterward, another timestamp is recorded when the anomaly is stopped. The database then resumes normal operation for a while before the next injection. This injection cycle is repeated, where each cycle introduces a randomly selected anomaly, ensuring the database cluster alternates between normal and abnormal states in a normal–abnormal–normal–abnormal... sequence.

\subsection{Evaluation Metrics}

In this paper, we adopt commonly used evaluation metrics—Macro Precision, Macro Recall, and Macro F1-Score—to assess the performance of failure diagnosis models. These metrics are also widely employed in previous studies. Macro Precision, Macro Recall, and Macro F1-Score represent the arithmetic mean of the Precision, Recall, and F1-Score values across all classes, respectively.

Precision measures, for a given anomaly type, the proportion of correctly predicted instances among all instances predicted as that type. It is calculated as shown in Equation~\ref{eq: Precision}.

\begin{equation}
	P_i = \frac{TP_i}{TP_i + FP_i}
	\label{eq: Precision}
\end{equation}

Recall measures, for a given anomaly type, the proportion of correctly predicted instances among all actual instances of that type. It is calculated as shown in Equation~\ref{eq: Recall}.

\begin{equation}
	R_i = \frac{TP_i}{TP_i + FN_i}
	\label{eq: Recall}
\end{equation}

The F1-Score balances the influence of Precision and Recall and provides a comprehensive evaluation of the classifier's performance. It is defined as the harmonic mean of Precision and Recall, as shown in Equation~\ref{eq: F1}.

\begin{equation}
	F_i = \frac{2 \times P_i \times R_i}{P_i + R_i}
	\label{eq: F1}
\end{equation}

Consequently, the calculations for Macro Precision, Macro Recall, and Macro F1-Score are defined in Equation~\ref{eq: macro}.

\begin{equation}
	\begin{cases}
		MP = \frac{1}{N} \sum_{i=1}^{N} P_i \\
		MR = \frac{1}{N} \sum_{i=1}^{N} R_i \\
		MF = \frac{1}{N} \sum_{i=1}^{N} F_i
	\end{cases}
	\label{eq: macro}
\end{equation}

\subsection{Experiment Setup}

We compare the proposed method, LogDB, with existing log-based failure diagnosis approaches, including LogKG~\cite{sui2023logkg}, Cloud19~\cite{yuan2019approach}, and LogCluster~\cite{lin2016log}. LogKG~\cite{sui2023logkg} employs a fault-oriented log representation and OPTICS clustering to extract entities and relationships from log data. LogCluster~\cite{lin2016log} clusters log messages and cross-references them with a knowledge base. Cloud19~\cite{yuan2019approach} utilizes word2vec to extract representation vectors from anomalous logs for fault type classification. Since these methods were originally designed for single-node failure diagnosis—specifically, classifying the fault type on the node where the anomaly was injected—we follow this setting by performing classification only on the injected node and calculating the corresponding evaluation metrics.

The choice of experimental parameters can affect the results. In the LogDB failure diagnosis workflow, four key parameters need to be preset: the time window size $T$, the log window size $N$, the length of the input feature matrix $\beta$ for the variational autoencoder, and the size of the latent matrix $\theta$. Unless otherwise specified, we set the time window size to 5 seconds, the log window size to 100, the input feature matrix length to 64, and the latent matrix size to 16.

Furthermore, to comprehensively evaluate different failure diagnosis methods, we also perform classification under normal system operation, labeling this category as No.0.

\subsection{Overall Evaluation}

\begingroup
\setlength{\tabcolsep}{15pt}
\begin{table}[htbp]
	\centering
	\caption{Performance Comparison of LogDB with Existing Methods}
	\label{tab: overall-evaluation}
	\begin{tabular}{ccccc}
		\toprule
		Model &  & TSBS & TPCx-IoT & IoT-Bench\\
		\midrule
		\multirow{3}*{LogKG} & \textbf{P} & 88.52\% & 50.19\% & 50.71\%\\
		~ & \textbf{R} & 76.78\% & 52.98\% & 53.01\%\\
		~ & \textbf{F1} & 81.01\% & 48.25\% & 50.61\%\\
		\midrule
		\multirow{3}*{LogCluster} & \textbf{P} & 71.34\% & 43.75\% & 25.93\%\\
		~ & \textbf{R} & 66.16\% & 50.00\% & 33.33\%\\
		~ & \textbf{F1} & 66.33\% & 46.67\% & 29.17\%\\
		\midrule
		\multirow{3}*{Cloud19} & \textbf{P} & 90.46\% & 78.53\% & 89.61\%\\
		~ & \textbf{R} & 71.43\% & 71.57\% & 77.35\%\\
		~ & \textbf{F1} & 78.68\% & 68.93\% & 82.22\%\\
		\midrule
		\multirow{3}*{\textbf{LogDB}} &  \textbf{P} & 99.72\% & 95.14\% & 89.61\%\\
		~ & \textbf{R} & 98.21\% & 97.03\% & 87.04\%\\
		~ & \textbf{F1} & 98.06\% & 95.76\% & 87.62\% \\
		\bottomrule
	\end{tabular}
\end{table}
\endgroup

To assess the effectiveness of LogDB in log-based failure diagnosis for distributed databases, we compare it against several state-of-the-art log-based approaches.

As shown in Table~\ref{tab: overall-evaluation}, our method significantly outperforms the graph-based diagnosis methods LogKG and LogCluster, achieving the best results across all three benchmark workloads: TSBS, TPCx-IoT, and IoT-Benchmark. In particular, LogKG and LogCluster fail to achieve accurate classification under TPCx-IoT and IoT-Benchmark workloads. This is because databases are inherently complex and dynamic systems composed of numerous interrelated and evolving components. Graph-based models struggle to capture these internal complexities and dynamic behaviors, especially when workload patterns change.

Compared with the deep learning-based method Cloud19, LogDB improves F1-score by 19.38\%, 26.83\%, and 5.40\% on TSBS, TPCx-IoT, and IoT-Benchmark workloads, respectively. Although Cloud19 performs better than graph-based methods by leveraging learned log embeddings, it only classifies anomalies based on logs from a single database node, which often leads to misjudgment. In contrast, LogDB aggregates logs from all database nodes, enabling a more holistic and accurate diagnosis of cluster-wide anomalies.

Overall, LogDB demonstrates clear advantages over most existing methods, offering stronger diagnostic guidance and higher practical value.

\subsection{Class-Wise Study}

\begingroup
\setlength{\tabcolsep}{12pt}
\begin{table}[htbp]
	\centering
	\caption{Diagnosis Performance of Each Anomaly Type}
	\label{tab: class-wise-evaluation}
	\begin{tabular}{cccccc}
		\toprule
		Anomaly &  & \textbf{LogKG} & \textbf{LogCluster} & \textbf{Cloud19} & \textbf{LogDB}(\textit{ours})\\
		\midrule
		\multirow{3}*{No.0} & \textbf{P} & 96.30\% & 83.10\% & 89.96\% & 97.73\%\\
		~ & \textbf{R} & 99.43\% & 91.76\% & 99.30\% & 100.00\%\\
		~ & \textbf{F1} & 97.84\% & 87.21\% & 94.40\% & 98.85\%\\
		\midrule
		\multirow{3}*{No.1} & \textbf{P} & 98.71\% & 98.86\% & 93.07\% & 100.00\%\\
		~ & \textbf{R} & 74.22\% & 82.94\% & 65.12\% & 100.00\%\\
		~ & \textbf{F1} & 84.73\% & 90.20\% & 76.62\% & 100.00\%\\
		\midrule
		\multirow{3}*{No.2} & \textbf{P} & 71.61\% & 15.48\% & 94.51\% & 100.00\%\\
		~ & \textbf{R} & 92.39\% & 36.27\% & 73.72\% & 100.00\%\\
		~ & \textbf{F1} & 80.69\% & 21.70\% & 82.83\% & 100.00\%\\
		\midrule
		\multirow{3}*{No.3} & \textbf{P} & 68.30\% & 50.14\% & 85.30\% & 100.00\%\\
		~ & \textbf{R} & 66.48\% & 13.75\% & 41.25\% & 100.00\%\\
		~ & \textbf{F1} & 67.38\% & 21.58\% & 55.61\% & 100.00\%\\
		\midrule
		\multirow{3}*{No.4} & \textbf{P} & 92.45\% & 97.64\% & 87.27\% & 100.00\%\\
		~ & \textbf{R} & 66.48\% & 13.75\% & 41.25\% & 100.00\%\\
		~ & \textbf{F1} & 61.17\% & 93.20\% & 54.66\% & 93.33\%\\
		\midrule
		\multirow{3}*{No.5} & \textbf{P} & 97.95\% & 72.00\% & 93.91\% & 100.00\%\\
		~ & \textbf{R} & 67.15\% & 95.38\% & 65.37\% & 85.71\%\\
		~ & \textbf{F1} & 85.69\% & 82.05\% & 77.08\% & 92.31\%\\
		\midrule
		\multirow{3}*{No.6} & \textbf{P} & 89.66\% & 75.39\% & 92.87\% & 100.00\%\\
		~ & \textbf{R} & 79.95\% & 60.02\% & 91.91\% & 100.00\%\\
		~ & \textbf{F1} & 84.53\% & 66.84\% & 92.39\% & 100.00\%\\
		\midrule
		\multirow{3}*{No.7} & \textbf{P} & 93.19\%	& 78.09\% & 96.81\% & 100.00\%\\
		~ & \textbf{R} & 79.89\% & 60.05\% & 94.97\% & 100.00\%\\
		~ & \textbf{F1} & 86.03\% & 67.89\% & 95.88\% & 100.00\%\\
		\bottomrule
	\end{tabular}
\end{table}
\endgroup

To further evaluate the failure diagnosis capabilities of LogDB, we recorded the diagnostic performance of LogDB and several baseline methods across different anomaly types, as shown in Table~\ref{tab: class-wise-evaluation}.

We observe that the diagnostic performance varies significantly across different anomaly types. For anomalies such as No.0 (Normal), No.1 (CPU Anomaly), No.6 (Background Task Anomaly), and No.7 (Excessive Flushing Speed), all methods are generally able to achieve effective identification.

However, for anomalies like No.2 (Memory Anomaly) and No.3 (Network Anomaly), LogCluster almost fails to recognize them. This is because memory and network anomalies often cause a sharp decrease in the number of logs generated during the anomalous period, rendering LogCluster’s clustering-based method ineffective.

For anomalies such as No.4 (Bulk Data Export Anomaly) and No.5 (Bulk Data Import Anomaly), both LogKG and Cloud19 show relatively poor performance. The reason is that these types of anomalies typically involve only a small number of related logs on a single node, requiring cross-node log information for accurate identification. Moreover, because the log patterns of these anomalies differ only slightly from those during normal operation, even LogDB may incur a certain level of false positives.

Nevertheless, apart from these specific cases, LogDB consistently achieves strong diagnostic results across other anomaly types.

Overall, compared with existing baseline methods, LogDB significantly improves diagnostic performance across all types of anomalies.

\subsection{Hyperparameter Analysis}

To thoroughly investigate how our algorithm is influenced by different model and input parameters, we conducted a series of experiments evaluating LogDB's performance under varying autoencoder and input settings. Specifically, we examined the effects of varying the input feature matrix length, the size of the latent matrix in the variational autoencoder, the number of nodes whose logs are used, and the size of the time window on LogDB's failure diagnosis performance.

\begin{figure}[htbp]
	\centering
	\subfigure[Input Feature Matrix Length]{
		\begin{minipage}{0.48\linewidth}
			\centering   
			\includegraphics[width=\textwidth]{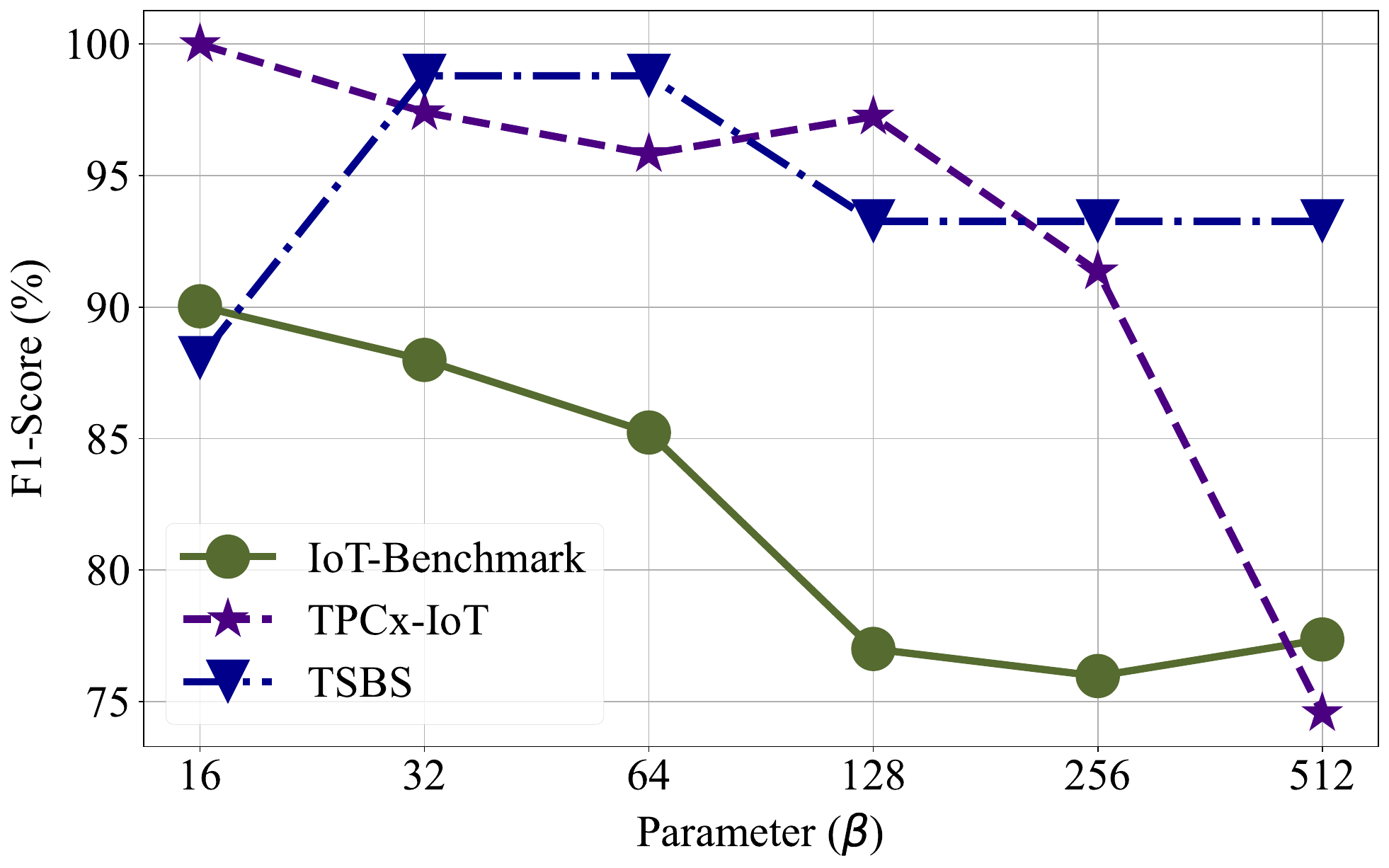}
			\label{fig: parameter-input-dim}
		\end{minipage}
	}
	\subfigure[Latent Matrix Size]{
		\begin{minipage}{0.48\linewidth}
			\centering
			\includegraphics[width=\textwidth]{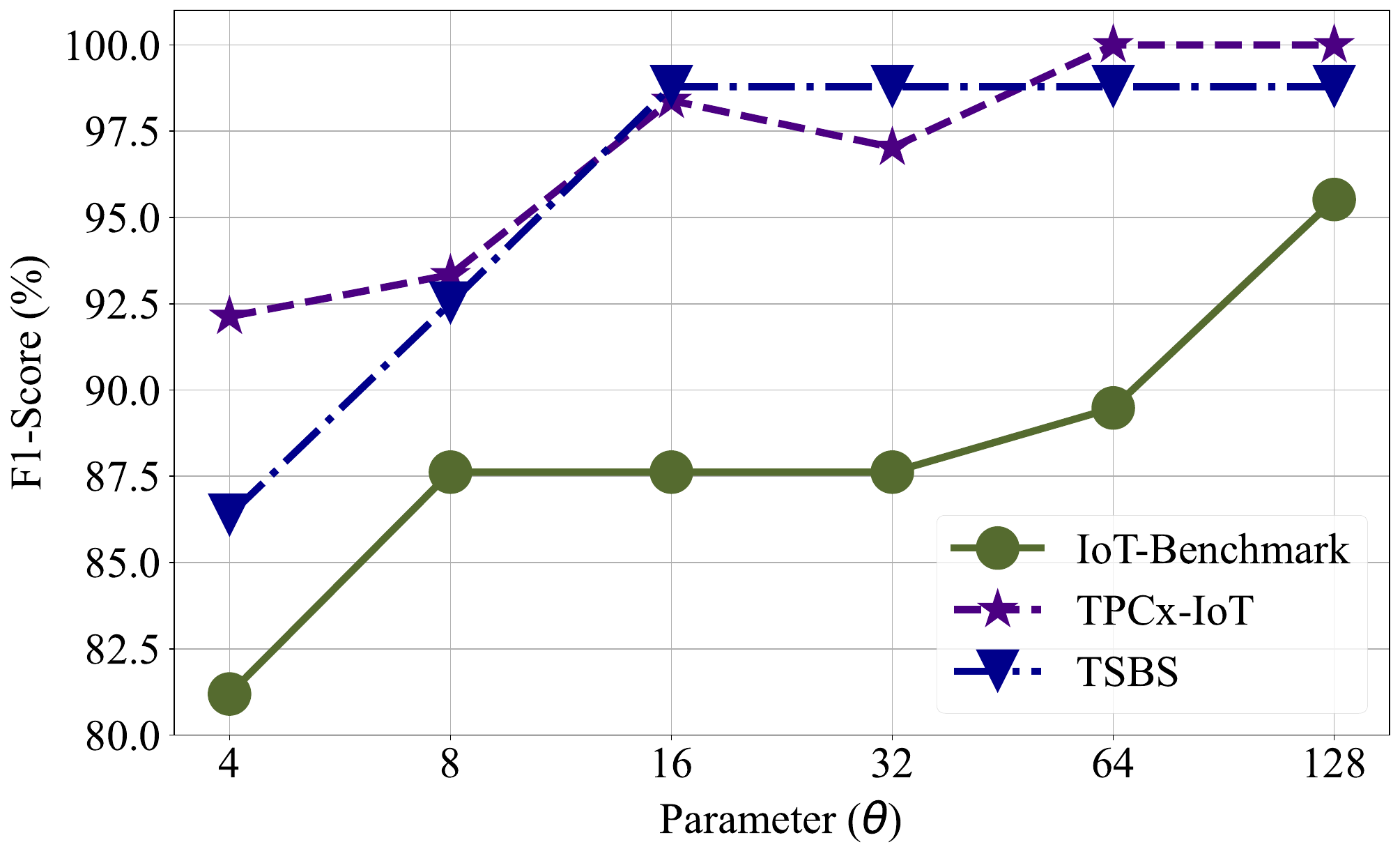}
			\label{fig: parameter-output-dim}
		\end{minipage}
	}
	\caption{Performance of LogDB under different autoencoder parameters.}
	\label{fig: parameter-experiment}
\end{figure}

Figure~\ref{fig: parameter-input-dim} shows the performance of LogDB under different input feature matrix lengths. As observed, the failure diagnosis performance first improves and then declines as the matrix length increases. This trend arises because a shorter input matrix fails to capture sufficient system features from different database nodes, impairing accurate failure diagnosis. Conversely, an overly long matrix retains excessive noise, degrading the classification performance. Nonetheless, across all matrix lengths, LogDB maintains a relatively high and stable diagnostic accuracy.

The influence of the latent matrix size on diagnostic performance varies under different workloads. As illustrated in Figure~\ref{fig: parameter-output-dim}, for the TPCx-IoT and TSBS workloads, increasing the latent matrix size improves the model performance up to a certain point, beyond which performance stabilizes. In contrast, for the IoT-Benchmark workload, increasing the latent matrix size consistently enhances performance. This is attributed to the complexity of the IoT-Benchmark workload, which embeds richer implicit information in the system logs.

\begin{figure}[htbp]
	\centering
	\subfigure[Number of Nodes Used]{
		\begin{minipage}{0.48\linewidth}
			\centering   
			\includegraphics[width=\textwidth]{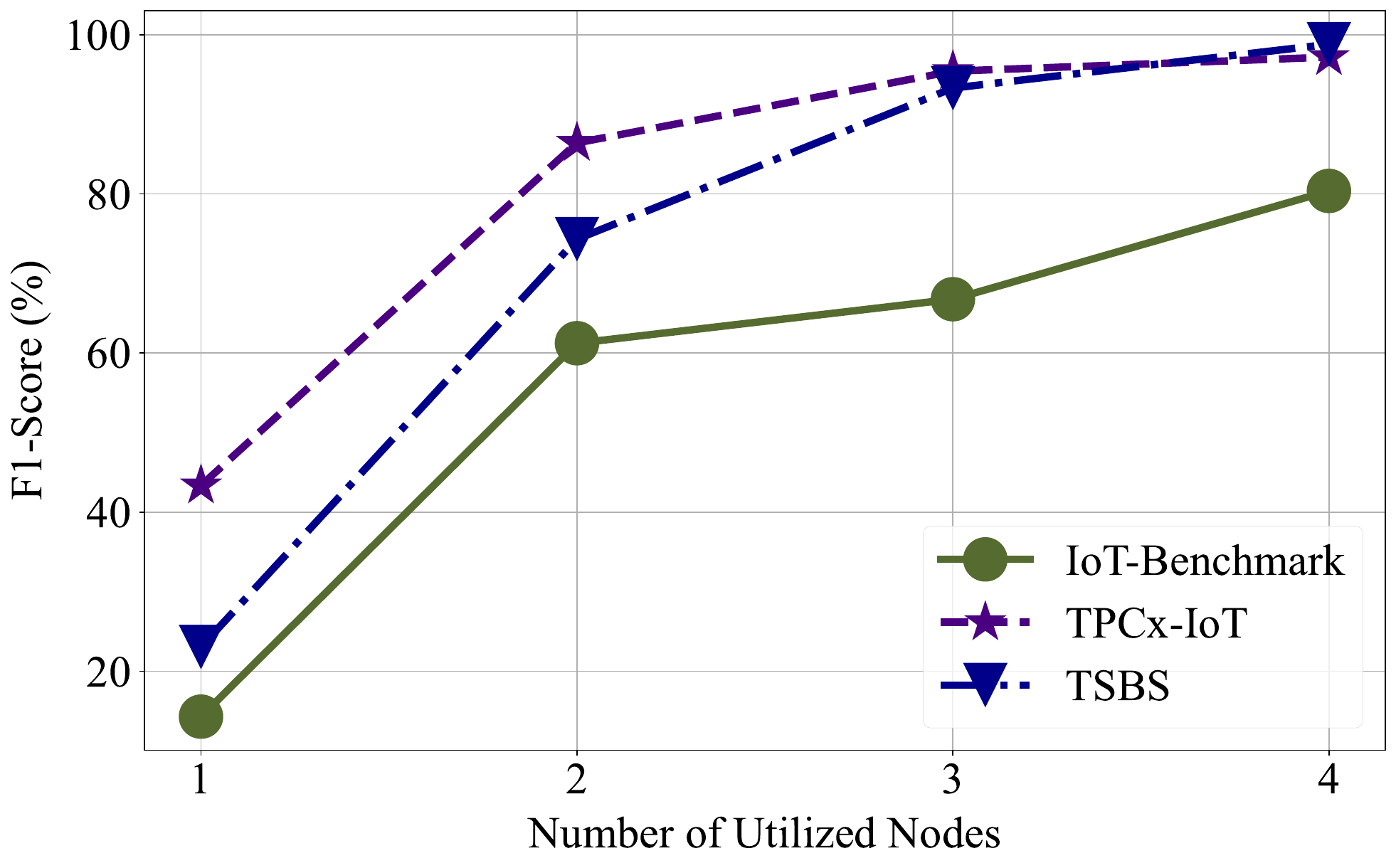}
			\label{fig: parameter-use-node}
		\end{minipage}
	}
	\subfigure[Time Window Size]{
		\begin{minipage}{0.48\linewidth}
			\centering
			\includegraphics[width=\textwidth]{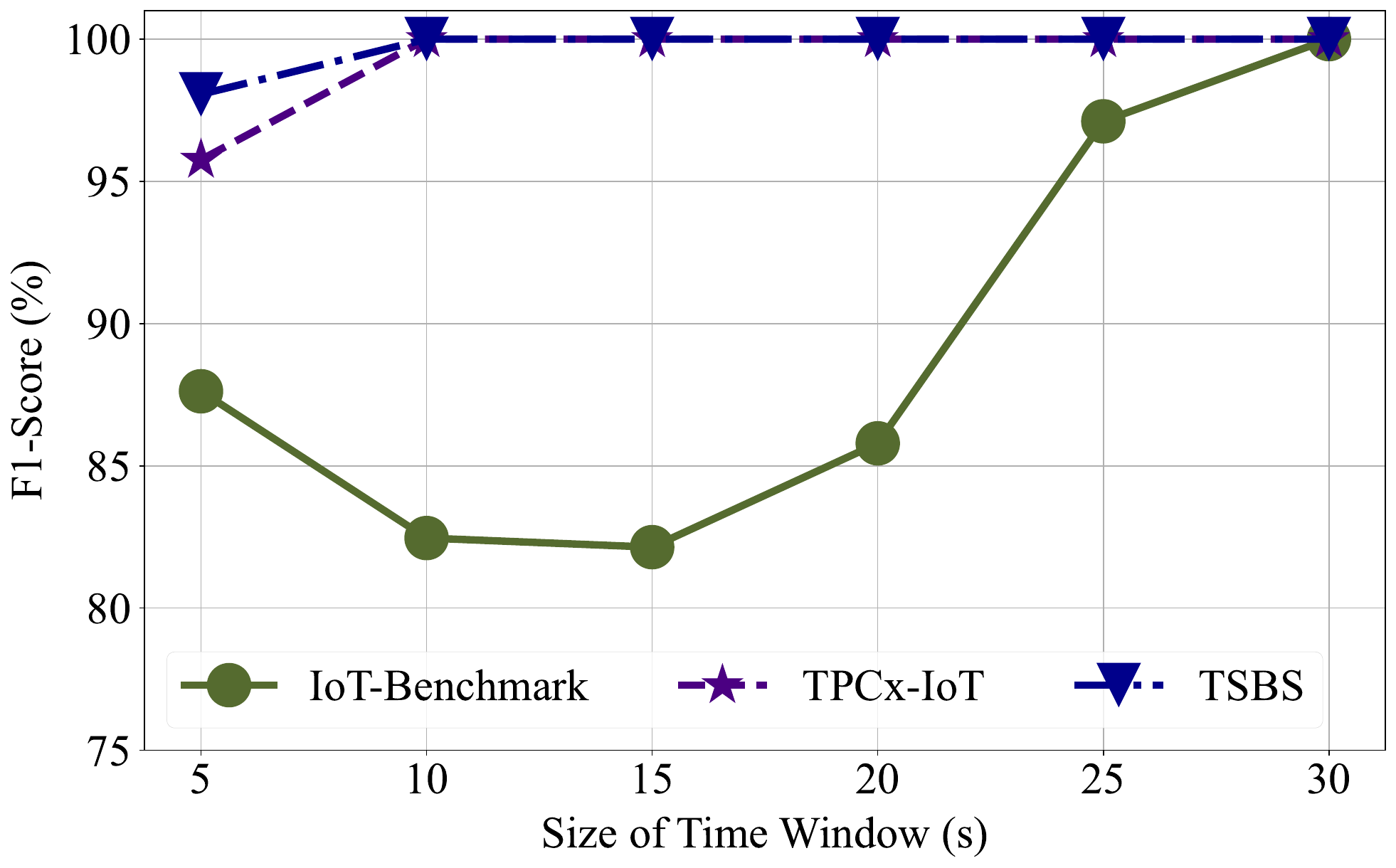}
			\label{fig: parameter-window}
		\end{minipage}
	}
	\caption{Performance of LogDB under different input settings.}
	\label{fig: parameter-experiment2}
\end{figure}

The number of nodes whose logs are used significantly impacts LogDB's diagnostic performance. As shown in Figure~\ref{fig: parameter-use-node}, across all workloads, using logs from fewer nodes results in inferior diagnosis performance. For instance, under the TPCx-IoT workload, using logs from only a single node yields an F1-Score of merely 43.34\%. As the number of nodes increases, the failure diagnosis performance steadily improves, reaching its peak when logs from all four nodes are utilized.

Finally, Figure~\ref{fig: parameter-window} presents LogDB's performance under different time window sizes. For the TPCx-IoT and TSBS workloads, a larger time window consistently leads to better performance, even achieving 100\% F1-Score. This is because larger windows contain richer internal log information and allow for coarser-grained diagnosis. However, for the IoT-Benchmark workload, the performance first deteriorates and then sharply improves with larger windows. This phenomenon is due to the workload's complexity: moderately increasing the window size initially introduces excessive noise, hindering diagnosis, whereas much larger windows enable sufficiently coarse-grained analysis to overcome noise and improve model effectiveness. Overall, while LogDB achieves excellent performance with large time windows, in practical deployment, we prefer smaller windows to enable faster anomaly detection and prompt repair by system administrators.

In summary, LogDB consistently achieves strong failure diagnosis performance across various configurations, except under certain extreme conditions. These exceptions include scenarios where the latent matrix is excessively small or logs from too few nodes are used, both of which result in insufficient information being captured and thus degrade overall performance.

\section{Related Work}

\subsection{Failure Diagnosis For Database}

There have been extensive studies on failure diagnosis methods for databases. Depending on the type of data utilized and the specific diagnosis targets, existing works can be broadly categorized into three groups: (1) diagnosing the overall operational status of the database, (2) diagnosing SQL-related faults, and (3) diagnosing failures within internal database components.

Diagnosis of the overall database operational status focuses on identifying and classifying coarse-grained root causes~\cite{li2021opengauss,zhou2021dbmind,yoon2016dbsherlock,benoit2005automatic,liu2020fluxinfer,ji2023perfce}. These studies typically formulate the problem as a multi-class classification task, where the input consists of runtime system data and the output is the corresponding anomaly category, such as high CPU usage, excessive memory consumption, or network transmission issues. For example, openGauss-AI~\cite{li2021opengauss} and openGauss-DBMind~\cite{zhou2021dbmind} propose a complete pipeline including an LSTM-based time-series anomaly detection model, Kolmogorov-Smirnov test-based anomaly metric extraction, and an optimized K-nearest neighbors (KNN)-based root cause analysis method. DBSherlock~\cite{yoon2016dbsherlock} presents a performance diagnosis tool that leverages monitoring metrics to effectively address runtime performance anomalies. Benoit et al.\cite{benoit2005automatic} unify diagnosis and tuning tasks within a single decision tree framework. FluxInfer\cite{liu2020fluxinfer} constructs a weighted undirected dependency graph based on monitoring metrics to locate root causes.

Diagnosis targeting SQL-related faults aims to identify and classify fine-grained root causes at the SQL level~\cite{li2021opengauss,zhou2021dbmind,yoon2016dbsherlock,remil2021makes,dintyala2020sqlcheck,dundjerski2020automatic}. Similar to the overall system diagnosis, this is also framed as a multi-class classification problem. However, the targets are more fine-grained and query-centric, such as excessive indexing, heavy query load, or resource contention due to background tasks. Notably, after identifying a system-level anomaly as slow SQL, openGauss-AI~\cite{li2021opengauss} and openGauss-DBMind~\cite{zhou2021dbmind} further perform SQL-level diagnosis: openGauss-AI identifies costly SQL operators without actually running the SQL statements, while openGauss-DBMind employs LSTM models to trace the root causes of slow queries. iSQUAD~\cite{ma2020diagnosing} uses topic modeling and Naive Bayes classification to diagnose intermittent slow query anomalies. DIADS~\cite{remil2021makes} is tailored for DBMS systems built atop Storage Area Networks (SANs). Upon detecting abnormal query execution times, DIADS extracts operator-level metrics and leverages machine learning to identify problematic operators, ultimately diagnosing the underlying hardware causes via an annotated query plan graph.

Given the complex internal architecture of database systems, there is also a line of work dedicated to diagnosing failures within internal components~\cite{dias2005automatic,kalmegh2019iqcar,glasbergen2020sentinel,nagaraj2012structured,seo2022nosql,dexter2020detecting}. For instance, ADDM~\cite{dias2005automatic} introduces the concept of database time, measuring runtime at the granularity of resource modules and SQL execution phases, and constructs a DBTime graph for failure localization. iQCAR~\cite{kalmegh2019iqcar} focuses on analyzing concurrency control, investigating the impact of concurrent queries during execution. Sentinel~\cite{glasbergen2020sentinel} builds fine-grained system behavior models from debug logs to assist DBAs with diagnosis. Distalyzer~\cite{nagaraj2012structured} applies machine learning to correlate behaviors extracted from logs with performance anomalies, automatically inferring the strongest system component-performance relationships.

Our work falls under the category of diagnosing the overall operational status of distributed databases. To the best of our knowledge, most existing studies in this area primarily leverage monitoring metrics for diagnosis, whereas our approach is the first to perform database failure diagnosis based solely on log data.

\subsection{Log-Based Failure Diagnosis}

Log-based anomaly detection and diagnosis is a well-established field~\cite{sui2023logkg,lin2016log,yuan2019approach, jia2022augmenting,jia2021logflash,du2017deeplog,meng2019loganomaly,zhang2019robust,yang2021semi,liu2022uniparser}. These methods generally consist of two stages: first, extracting templates and key information from raw logs, and second, building models for anomaly detection or classification. Based on the underlying modeling approach, existing methods can be categorized into two major groups: graph-based methods and deep learning-based methods.

Graph-based methods construct graph representations of system behaviors using parsed log events. Anomalies are diagnosed by detecting conflicts between the runtime event sequence and the established graph. For example, LogFlash~\cite{jia2021logflash} employs real-time stream processing for efficient log transformation, improving detection speed. HiLog~\cite{jia2022augmenting} investigates anti-patterns that challenge common assumptions in log-based anomaly detection and proposes a human-in-the-loop approach to integrate domain knowledge. LogCluster~\cite{yuan2019approach} clusters logs to simplify the log model and uses a knowledge graph for diagnosis. LogKG~\cite{sui2023logkg} introduces a fault-oriented log representation (FOLR) to extract fault-related patterns and uses OPTICS clustering for anomaly detection. SwissLog~\cite{li2022swisslog} tackles the challenge of interleaved logs generated by concurrent microservice components by building cross-component ID graphs for anomaly detection.

Deep learning-based methods employ neural networks to model sequential patterns in log data. DeepLog~\cite{du2017deeplog} uses LSTM networks to learn both the sequence of log events and the variables embedded within log texts. LogAnomaly~\cite{meng2019loganomaly} utilizes word2vec embeddings to enrich the semantic representation of events. LogRobust~\cite{zhang2019robust} applies TF-IDF and word embeddings to transform logs into semantic vectors. UniParser~\cite{liu2022uniparser} introduces a token encoder and a context encoder to learn patterns from log tokens and their surrounding contexts. Cloud19~\cite{yuan2019approach} focuses on identifying anomalous system task logs in cloud environments and matching new abnormal behavior patterns against known ones for diagnosis.

In summary, graph-based methods typically offer faster runtime efficiency, as they only require constructing and comparing log graphs during the training and inference phases. However, they often struggle to learn rich latent representations, resulting in slightly lower detection and diagnosis accuracy compared to deep learning-based methods. In contrast, deep learning-based methods require longer training times but can better capture complex semantic patterns in logs. In this paper, we adopt a deep learning-based approach and compare its performance against both graph-based methods (e.g., LogKG, LogCluster) and deep learning-based methods (e.g., Cloud19) in our experiments.

\section{Conclusion}

failure diagnosis methods tailored for database systems can effectively classify and diagnose anomalies occurring within these systems, thereby assisting operators in timely system recovery. These methods are highly practical. Logs, as naturally occurring data within systems, provide valuable insights into the internal states of databases. However, existing log-based failure diagnosis methods are not specifically designed for database systems and often fail to fully leverage internal database characteristics and distributed features.

In this paper, we propose \textbf{LogDB}, a log-based failure diagnosis method tailored for distributed databases. To the best of our knowledge, LogDB is the first approach to perform fault diagnosis for database systems using log data. It collects and encodes sequential, quantitative, and semantic information from each independent node in a database cluster. These features are processed using an LSTM model with self-attention to generate node-specific anomaly feature matrices. A variational autoencoder is then employed on the primary node to fuse and extract implicit features from these matrices across nodes, ultimately enabling accurate failure diagnosis at the cluster level.

We validate the effectiveness of LogDB through extensive experiments on the distributed database \textit{Apache IoTDB}. Our results show that LogDB can effectively extract anomalous features from massive volumes of log data and integrate feature information across different nodes, achieving accurate classification of anomaly types at the cluster level.

Currently, LogDB has been implemented only on Apache IoTDB. Due to the lack of public benchmark datasets, the anomalies used in our experiments are generated through chaos testing and fault injection. As future work, we plan to collaborate with major database vendors to explore more realistic deployment scenarios. Moreover, we aim to extend LogDB by integrating multimodal data---including monitoring metrics, log data, and trace information---for more comprehensive failure diagnosis.

\bibliographystyle{ACM-Reference-Format}
\bibliography{sample-base}

\end{document}